\documentclass[pre,amsmath]{revtex4-2}
\usepackage{graphicx}
\usepackage{color}
\usepackage{bm}
\usepackage{epsfig}
\usepackage{latexsym}
\usepackage{nameref,hyperref}
\begin{document}
\newcommand{\be}{\begin{equation}}
\newcommand{\ee}{\end{equation}}
\newcommand{\bea}{\begin{eqnarray}}
\newcommand{\eea}{\end{eqnarray}}
\newcommand{\RN}[1]{\textup{\uppercase\expandafter{\romannumeral#1}}}
\title{Short time extremal response to step stimulus for a single cell  {\sl {E. coli}}}
\author{Sakuntala Chatterjee}
\affiliation{Department of Physics of Complex Systems, S. N. Bose National Centre for Basic Sciences, Block JD, Sector 3, Salt Lake, Kolkata 700106, India.}
\begin{abstract}
After application of a step stimulus, in the form of a sudden change in attractant environment, the receptor activity and tumbling bias of an  {\sl E. coli} cell change sharply to reach their extremal values before they gradually relax to their post-stimulus adapted levels in the long time limit. We perform numerical simulations and exact calculations to investigate the short time response of the cell. For both activity and tumbling bias, we exactly derive the condition for extremal response and find good agreement with simulations. We also make experimentally verifiable prediction that there is an optimum size of the step stimulus at which the extremal response is reached in the shortest possible time.  
\end{abstract}
\maketitle

\section{Introduction}

Chemotaxis refers to directed motion in response to a chemical signal \cite{eisenbachbook}. An  {\sl E. coli} cell uses its transmembrane receptors to sense the chemical environment and regulates its run-and-tumble motility to navigate \cite{berg2008coli, tu2013quantitative, colin2017emergent, lan2016information}. The intracellular signaling pathway that allows the cell to migrate towards more favorable environment is one of the most well-characterized systems in biology, both experimentally and theoretically \cite{berg1972chemotaxis, alon1999robustness,  keymer2006chemosensing, shimizu2010modular, tu2008modeling, bi2018stimulus, kafri2008steady, chatterjee2011chemotaxis}. How chemotactic behavior of the cell is affected by the diversity present across the cell population, or even by noise present within a single cell has received a lot of research attention lately \cite{emonet2008relationship, keegstra2017phenotypic, colin2017multiple, frankel2014adaptability, flores2012signaling, mello2020sequential, aquino2011optimal, dev2018optimal, shobhan, dev2015search, dev2019rnt, kamino2020adaptive, gumerov2021diversity, colin2021multiple}.

The transmembrane chemoreceptors bind and unbind to the attractant molecules in the extra-cellular environment and send a phosphorylation-based signal to the flagellar motors to control the run-and-tumble motion \cite{studdert2005insights, pontius2013adaptation, parkinson2015signaling}. Based on their phosphorylation activity, the chemoreceptors are considered to be in two different states, active and inactive \cite{tu2008modeling, jiang2010quantitative, keymer2006chemosensing}. In the active state the receptors promote phosphorylation which in turn induces clockwise rotation of the flagellar motors and the cell tumbles. In most experiments either the chemoreceptor activity, or the motor bias is measured as the chemotactic response of the cell \cite{block1982impulse, keegstra2017phenotypic, colin2017multiple, min2012chemotactic}. Following a stimulus signal in the form of a sudden change in attractant environment, these quantities undergo sharp change in the short time limit, followed by slow recovery in the long time limit, controlled by the adaptation module of the signaling network \cite{min2012chemotactic, frank2013prolonged, colin2017multiple, keegstra2017phenotypic}. Many experimental and theoretical studies in literature focus on this long time relaxation which gives information about the adaptation time-scale and how perfect or robust the adaptation process is. The early time sharp response is often interpreted as `chemotactic sensitivity' and larger the magnitude of deviation from pre-stimulus level, higher is considered the sensitivity \cite{sourjik2002receptor, waite2018behavioral, tu2013quantitative}. However, very little is known about the dynamics of the short time extremal response. How long does the cell need to reach its extremal response and how to characterize the properties of the signaling network at this time? In this work, we focus on these issues.

We use a theoretical model to describe the intracellular signaling network and run-and-tumble motility of the cell. Using extensive numerical simulations we measure the temporal variation of receptor activity and tumbling bias after the application of a step stimulus. We consider the case of step addition (removal) of attractant, {\sl i.e.} when the attractant level in the environment is suddenly increased (decreased) at time $t=0$ and then held at that elevated (reduced) level. In the pre-stimulus state $t<0$ and in the long time limit $t \to \infty$ the cell is expected to be in an adapted state with its environment and both activity and tumbling bias assume stationary values. However, these quantities show rapid variation with time immediately after the stimulus is given and reach extremal values at short times. We perform exact calculations to derive the extremal conditions for these quantities and find good agreement with simulations. Interestingly, these extremum conditions show some similarity with an adapted state, although the system is far from adaptation here.

Moreover, the time to reach the extremal response also shows interesting behavior for both activity and tumbling bias. The chemoreceptors have cooperative interaction among themselves which gives rise to receptor clustering \cite{maddock1993polar, pinas2016source}. This cooperativity amplifies the input signal and is the reason behind high sensitivity shown by the signaling network \cite{bray1998receptor, duke1999heightened, frank2016networked}. Our simulations show that there is an optimum size of the receptor cluster at which the cell reaches its extremal response in the shortest possible time. We explain the reason behind this interesting effect and we further argue that the same mechanism also yields an optimum step size of the stimulus at which the extremal response is reached fastest when the receptor cluster size is held fixed. Our numerical simulations verify this prediction. We also propose simple experiments to test our theory.

\section{Model description} 
The chemoreceptors form trimer of dimers (TDs) and  each receptor cluster of size $n$ contains $n$ such TDs \cite{briegel2012bacterial} \cite{liu2012molecular}. For simplicity, we assume all clusters to be of same size. For a total of $N_{dim}$ dimers in the cell, there are $N_{dim}/3n$ number of clusters or signaling teams in the system. According to Monod-Wyman-Changeux (MWC) model \cite{monod1965nature, changeux2005allosteric} all receptors within one team switch their activity state in unison. The free energy difference $F$ (in units of $k_B T$) between the active and inactive states of a cluster of size $n$ has the form: 
\be 
F = 3n \log \frac{1+[L]/K_{min}}{1+[L]/K_{max}} + 3n \epsilon_0 - \epsilon_1 \sum_{j=1}^{3n} m_j.   \label{eq:free}
\ee
Here, $[L]$ denotes the attractant concentration in the extra-cellular environment and $m_j$ denotes the methylation level of the $j$-th dimer in the cluster which can take any integer value between $0$ and $8$. We explain in the next paragraph how the enzyme molecules present in the cell modify $m_j$ through their binding and unbinding kinetics. In steady state when the cell is fully adapted to its attractant environment, the total activity of the cell, defined as the fraction of receptor clusters in the active state, is given by the Boltzmann distribution $\langle [1+e^F]^{-1} \rangle $ where the angular bracket denotes averaging over all methylation levels. For a given receptor cluster, $F$ controls the transition between the active and inactive states. In our model we have chosen the transition rates from local detailed balance. A receptor cluster with a total methylation level $m = \sum_{j=1}^{3n} m_j $ can switch from an inactive to active state with a rate proportional to $[1+e^F]^{-1} $ while the reverse transition rate goes as $[1+e^{-F}]^{-1}$ \cite{colin2017multiple, shobhan, mandal2022effect, shobhaninjp}. These rates are explicit functions of $m$.

Methylating enzyme CheR and phosphorylated demethylating enzyme CheB-P modify the value of $m_j$ upon binding with the dimer. In its inactive state, the dimer gets methylated by CheR which raises $m_j$ by $1$, provided $m_j < 8$. Similarly in active state, CheB-P can lower $m_j$ by $1$ provided $m_j > 0$. In Appendix \ref{app:model} we include detailed description of binding-unbinding kinetics of the enzyme molecules. From Eq. \ref{eq:free} it follows that methylation decreases free energy (hence increases activity) and demethylation does the opposite. This constitutes the negative feedback which is responsible for adaptation in the system. In an {\sl E. coli} cell there are only few hundred enzyme molecules, as opposed to few thousand receptors. To explain the near-perfect adaptation \cite{goy1977sensory, berg1975transient} shown by the cell, mechanisms like assistance neighborhood or brachiation were proposed in earlier works \cite{levin2002binding, endres2006precise, hansen2008chemotaxis, kim2002dynamic, li2005adaptational}. In an assistance neighborhood model, one enzyme molecule, while being tethered to one particular dimer, can also modify the methylation levels of the neighboring receptors. In brachiation model, the enzyme molecule, once bound to a dimer can perform random walk on the receptor array and move from one dimer to others and modify their methylation levels. We include a flavor of these mechanisms in our simple model, as explained in more details in Appendix \ref{app:model}. The total activity $a$ of the cell determines the rotational bias of the flagellar motors. When the motors are in CCW rotation mode, they can switch to CW mode with rate $\omega e^{-G(a)}$ and the reverse switch happens with rate $\omega e^{G(a)}$ where the function $G(a)$ has been described in Appendix \ref{app:model}. 

It follows from the discussion in the previous paragraph that the receptor activity acts as the connector between two principal modules of the chemotactic reaction network: sensing and adaptation. The sensing module responds to the stimulus signal by controling the run-and-tumble motion of the cell, while the adaptation module makes sure that the receptor activity does not get too high or too low. The schematic diagram of the reaction network presented in Fig. \ref{fig:model} makes this point clearer. The values of all model parameters are listed in Table \ref{table} in Appendix \ref{app:model}. 
\begin{figure}
\includegraphics[scale=1.2]{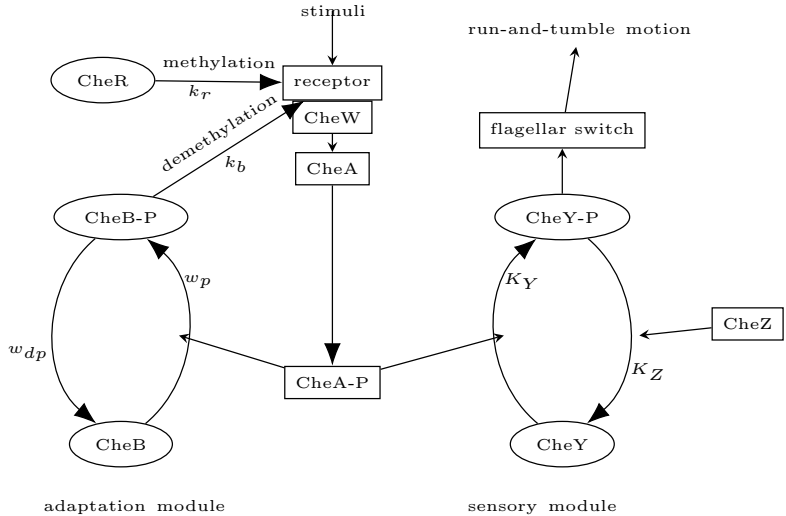}
\caption{Simplified schematic diagram of the signaling network inside an {\sl E. coli} cell.  The rates of (de)phosphorylation reactions and (de)methylation reactions are also shown.}
\label{fig:model}
\end{figure}

\section{Extremal value for receptor activity} 

\begin{figure}[h!]
\includegraphics[scale=0.7]{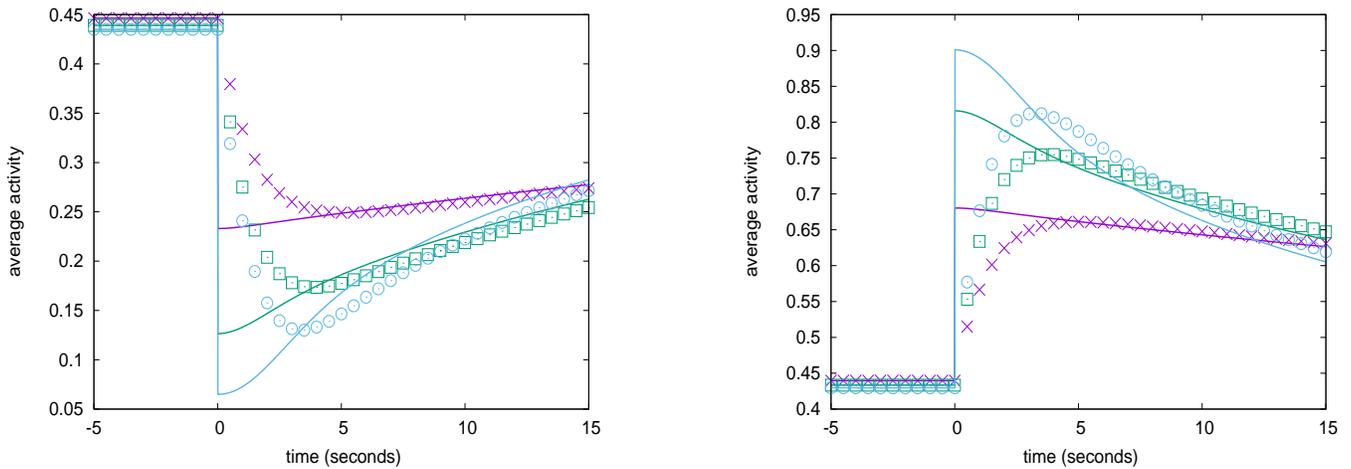}
\caption{Plot of average activity $\langle a(t) \rangle $ (discrete symbols) and $\langle [1+e^{F(t)} ]^{-1} \rangle $ (continuous lines) as a function of time. The left panel corresponds to step addition of stimulus with $[L]_0=200 \mu M$ and $[L]_1 = 250 \mu M$, while the right panel shows data for step removal with $[L]_0=300 \mu M$ and $[L]_1 = 200 \mu M$. Purple (cross) , green (squares) and blue (circles) colors correspond to different $n=5,10,15$, respectively. For all cases, at $t=t_a$ when the average activity hits its extremum level, $\langle a(t) \rangle $ and $\langle [1+e^{F(t)} ]^{-1} \rangle $ coincide. All other simulation parameters are as in Table \ref{table} of Appendix \ref{app:model}. } 
\label{fig:db}
\end{figure}

We consider a pre-stimulus attractant level $[L]_0$ with which the cell is completely adapted (also see Fig. \ref{fig:a0}). At time $t=0$ the attractant level is changed from $[L]_0$ to $[L]_1$. From the expression for free energy given in Eq. \ref{eq:free} it follows that for step addition (removal) of attractant $F$ shows a sharp increase (drop) at $t=0$. At subsequent times the methylation levels change slowly such that $F$ varies slowly with time and relaxes to its new post-stimulus adapted value. $F([L]_1, m(t))$ is the free energy difference between the active and inactive states of a receptor cluster whose methylation level is $m$ at time $t$. A receptor cluster at time $t$ switches from inactive to active state with a rate that is proportional to $(1+exp[{F([L]_1, m(t))}] )^{-1} $. In our simulations we average this time-varying rate over all receptor clusters with different methylation levels and plot the resulting quantity $ \langle [1+e^{F(t)} ]^{-1} \rangle $ in Fig. \ref{fig:db} (continuous lines). Here, the angular brackets denote averaging over various receptor clusters and for notational simplicity, we have not explicitly written the $[L]_1$ dependence of $F$.

In Fig. \ref{fig:db} we also show, by discrete symbols, the temporal variation of the average activity $\langle a(t) \rangle $ of the cell. We count the number of active clusters at time $t$ and divide it by the total number of clusters and average over histories. Our data show that $\langle a(t) \rangle$ starts changing immediately after the step stimulus is applied and after a finite amount of time $t_a$ it reaches its extremal value. For $t \gg t_a$ average activity gradually recovers towards the post-stimulus adapted level. For the same size of the step stimulus, we find larger change in activity as the receptor cluster size $n$ increases. We have also found that for the step addition and step removal of stimulus, the extremal response of the system shows an asymmetry. Our data in Fig. \ref{fig:asym} left panel compares the magnitude of extremal response between two step stimuli when $[L]_0$ and $[L]_1$ values are interchanged. We observe higher magnitude for the case of step removal. Interestingly, $t_a$ takes higher values for the case of step addition (see Fig. \ref{fig:asym} right panel.)  Earlier studies \cite{meir2010precision} \cite{min2012chemotactic} have reported asymmetry in the long time response of the system ({\sl e.g.} recovery time-scale) when the sign of the step is reversed. We find even in the short time response of the system, this asymmetry is present. 

It is seen from our data in Fig. \ref{fig:db} that for $t<0$ in the pre-stimulus stage, $\langle a(t) \rangle$ matches with $ \langle [1+e^{F(t)} ]^{-1} \rangle $, as expected for an adapted system. However, as soon as the step stimulus is applied, these two quantities start differing and one expects them to become equal again for very large times (not shown in Fig. \ref{fig:db}) when the system has readapted to its new attractant environment. For small times when the system is far from adaptation, one does not expect the probability to find a receptor cluster in active state to follow Boltzmann distribution. But our data show at $t=t_a$, the average activity  $\langle a(t) \rangle$ becomes momentarily equal to  $\langle [1+e^{F(t)} ]^{-1} \rangle $. In the next subsection we perform an exact calculation to explain this effect. Note that this phenomenon is distinctly different from a response overshoot which has been reported earlier for {\sl E. coli} adaptation \cite{min2012chemotactic, zhang2018motor}, where the response function crosses the post-stimulus adapted value (which is a time-independent quantity) before finally settling down to that value. In the present case, $\langle a(t) \rangle$ intersects another time-varying quantity $\langle [1+e^{F(t)} ]^{-1} \rangle $ at short times, which is different from $\langle a(t) \rangle$ overshooting its final steady state value $a_\infty$.

\subsection{Exact calculation for extremal activity} 

Let $P(N_1,t)$ be the probability to find $N_1$ active receptor clusters at time $t$ after the stimulus has been applied. If $N$ is the total number of clusters, then $N_1$ can take any integer value between $0$ and $N$. For $N_1$ outside this range we assume $P(N_1,t) =0$ for all $t$.  The master equation governing time-evolution of $P(N_1,t)$ can be written as 
\bea
\nonumber
\frac{\partial P (N_1,t)}{\partial t} &=& \Gamma_{0 \rightarrow 1} (t) \left [ P(N_1 -1, t) (N-N_1 +1) - P(N_1,t) (N-N_1) \right ] \\
& +& \Gamma_{1 \rightarrow 0} (t) \left [ P(N_1+1,t) (N_1+1)-P(N_1,t) N_1 \right ]
\label{eq:pn1}
\eea
where we have defined  $\Gamma_{0 \rightarrow 1} (t)$ as the total transition rate from inactive to active state: $\Gamma_{0 \rightarrow 1} (t) = \sum_m \dfrac{w_a}{1+e^F} p(m,t|0)$ where the sum is over all possible methylation levels of a cluster. Here, $p(m,t|0)$ denotes the conditional probability that a cluster has total methylation level $m$, given that the cluster is inactive. As follows from Eq. \ref{eq:free}, the free energy $F$ depends on this total $m$. In a similar way one can define $\Gamma_{1 \rightarrow 0} (t) = \sum_m \dfrac{w_a e^F}{1+e^F} p(m,t|1)$. Multiplying Eqs. \ref{eq:pn1} by $N_1$ and summing over all $N_1$ allows for a large number of telescopic cancellations and we finally get
\be
\frac{d \langle N_1(t) \rangle}{dt} = \Gamma_{0 \rightarrow 1} (t) (N-\langle N_1(t) \rangle ) -  \Gamma_{1 \rightarrow 0} (t) \langle N_1(t) \rangle
\ee
The average activity $\langle a(t) \rangle $ is nothing but $\langle N_1(t) \rangle /N$ and hence left hand side of the above equation must vanish at $t=t_a$. This gives
\be 
\sum_m \frac{{\mathcal P}(m,t=t_a)}{1+e^F} = \sum_m p(m,1,t=t_a) 
\label{eq:dbproof}
\ee
where we have rewritten the conditional probabilities $p(m,t|0)$, $p(m,t|1)$ in terms of joint distributions $p(m,0,t)$, $p(m,1,t)$ and defined ${\mathcal P}(m,t) = p(m,0,t)+p(m,1,t)$. While  $p(m,1,t)$ denotes the joint probability to find an active receptor cluster with methylation level $m$ at time $t$, the quantity ${\mathcal P}(m,t)$ stands for the probability that a receptor cluster at time $t$ has methylation level $m$, irrespective of its activity state. The left hand side of Eq. \ref{eq:dbproof} is the expectation value of $[1+e^F]^{-1}$ calculated at $t=t_a$ and the right hand side is nothing but the probability to find a cluster in the active state, {\sl i.e.}
\be
\langle a(t) \rangle \Big |_{t=t_a} = \langle [1+e^{F(t)}]^{-1} \rangle \Big |_{t=t_a}.
\label{eq:actdb}
\ee
This explains our observation in Fig. \ref{fig:db}. Notice that on both sides of Eq. \ref{eq:actdb} the angular brackets indicate averaging over all receptor clusters, with different $m$ values. This averaging is crucial since no such equilibrium-like relationship holds at short times for each individual $m$ value. Only when a weighted average over different $m$ is considered, we have a simple relation like Eq. \ref{eq:actdb}. Indeed the time-evolution equation of the joint probability $p(m,1,t)$ would contain terms due to activity switching, as well as terms due to demethylation reactions (since active clusters only get demethylated). Only when summation over all $m$ values are performed, the terms corresponding to methylation dynamics cancel and we have Eq. \ref{eq:actdb}. Note that in our exact calculation above we have not used any specific time-scale for methylation. Our exact calculation remains valid for all time-scales. However, if we use the fact that methylation-demethylation reactions are much slower than activity switching events, then in the time-evolution equation for $p(m,1,t)$ we can neglect the demethylation terms which change $m$ and only retain the terms for activity switching. Thus, using time-scale separation, $\partial_t p(m,1,t)$ has a simpler form for each $m$. Averaging over different receptor clusters with different $m$ values and using the extremal condition at $t=t_a$ as before yields  Eq. \ref{eq:dbproof}.

A discussion about the choice of transition rates is in order here. In our model we have used the switching rate of a particular receptor cluster from an inactive to active state as $w_a/(1+e^F)$ and the rate of the reverse transition as $w_a e^F/(1+e^F)$. This specific choice is motivated from the switching dynamics for receptor activity used in \cite{colin2017multiple}. However, in principle, many other choices for the rates are possible which satisfy local detailed balance, {\sl i.e.} the ratio between forward and reverse transition equals $e^F$. Although even for these choices the extremal activity will still satisfy a relationship with free energy which is expected to hold only in adapted state, the specific form of this relationship may differ from Eq. \ref{eq:actdb}.

\subsection{ Optimum size of the step-stimulus for quickest extremal response}

\begin{figure}[h!]
\includegraphics[scale=0.7]{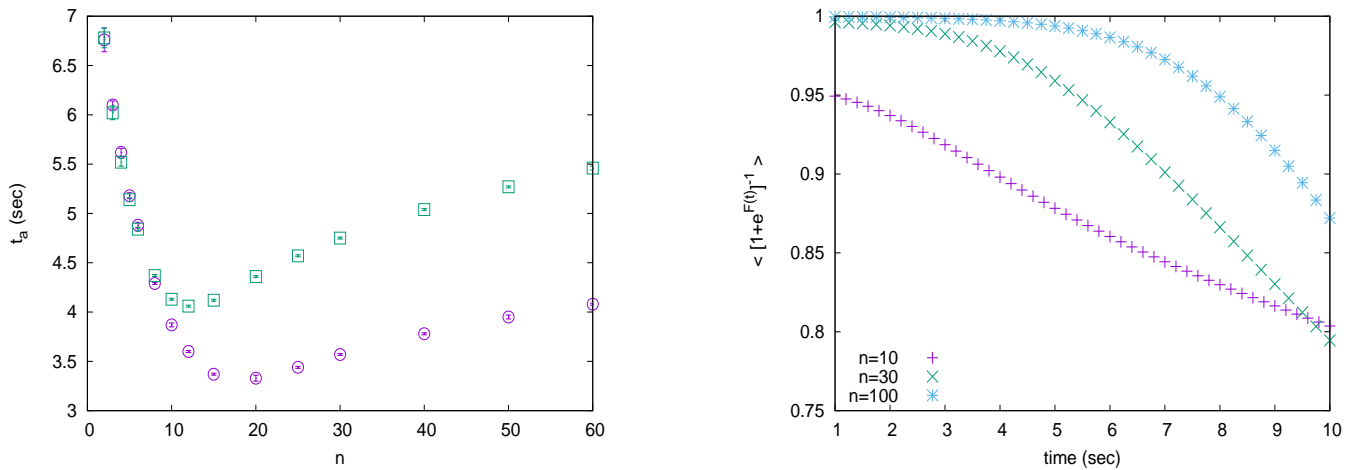}
\caption{Left panel: $t_a$ shows a minimum with $n$. The purple circles correspond to an increase in background attractant concentration from $200 \mu M$ to $250 \mu M$ and the green squares correspond to a decrease from $300 \mu M$ to $200 \mu M$. Error-bars are shown most of which are smaller than the symbol size. Right panel: For step removal of stimulus, the extremal value of $\langle [1+e^{F(t)}]^{-1} \rangle$ approaches unity for large $n$. It starts decreasing with time only after a significant amount of demethylation manages to lower the magnitude of $F$, which takes longer as $n$ increases. These data correspond to $[L]_0 = 300 \mu M$ and $[L]_1 = 200 \mu M$ and different $n$ values are shown in legends.} 
\label{fig:tm}
\end{figure}

To ensure that $t_a$ can be uniquely determined, we have considered only those ranges of step size or $n$ for which $\langle a(t) \rangle $ has a well-defined extremum. Since average activity can not be negative or larger than unity, the response of the system saturates when all clusters become active or inactive. The temporal variation of $\langle a(t) \rangle $ in that case shows a flat maximum or minimum from which $t_a$ can not be uniquely determined. To avoid this problem, we have here considered only those cases for which extremal value of activity remains strictly below $1$ and above $0$. We have verified that (Fig. \ref{fig:delact}) the difference between extremal activity and pre-stimulus activity increases linearly with the step size when the step size is small. The growth slows down for larger steps. In Fig. \ref{fig:tm} (left panel) we show the plot for $t_a$ with $n$ for step addition (purple circles) and step removal (green squares) of stimulus. Our data show in both cases $t_a$ shows a minimum with $n$. In other words, after the application of a step stimulus the system reaches its extremal response in the shortest possible time for a specific size of the receptor cluster. 

To explain this observation, let us define $z(t) = \langle [1+e^{F(t)}]^{-1} \rangle -\langle a(t) \rangle $, which can be interpreted as a measure of distance from the adapted state. $z(t)$ is the difference between two terms each of which is bounded between $0$ and $1$. Eq. \ref{eq:actdb} means that $z(t_a) =0$. It follows from Eq. \ref{eq:free} that the change in $F$ at $t=0$ scales with $n$. Consider first the case of step-removal of attractant. For large $n$, the decrease in $F$ is so large that $\langle [1+e^F]^{-1} \rangle$ reaches very close to unity, which effectively blocks all transitions from active to inactive state. Until the magnitude of $F$ is significantly reduced during subsequent modification of methylation levels, this transition remains blocked and the activity of the system can only increase. In this time range $z(t) \approx 1-\langle a(t) \rangle$ which can never become zero, since we have considered only those cases for which extremal activity is strictly below unity. As $n$ increases, it takes longer for the system to lower its methylation level enough such that $\langle [1+e^F]^{-1}\rangle$ falls below unity and active to inactive transition is possible again and activity can start decreasing. Our data in Fig. \ref{fig:tm} (right panel) clearly show this trend where we find that as $n$ takes larger values, $\langle [1+e^F]^{-1} \rangle$ remains stuck to unity for a longer time before it decreases again. This explains why $t_a$ increases with $n$ for large $n$.

For small $n$, the change in $F$ is also scaled down and  $\langle [1+e^{F(t)}]^{-1} \rangle$ remains significantly below unity. In this case, the active to inactive transition occurs with a smaller rate and the reverse transition occurs with a larger rate. Both these rates depend on $n$  and as $n$ increases (still remaining small), the difference between these two transition rates also goes up. Therefore activity increases faster and $z(t)$ reaches zero quicker. Although $z(t=0)$ has a higher positive value for larger $n$ (owing to steeper rise of $\langle [1+e^{F(t)}]^{-1} \rangle$), the faster increase of activity still manages to close the gap in a smaller time. This explains why for small $n$ we find $t_a$ decreases with $n$. Decreasing trend for small $n$ and increasing trend for large $n$ gives rise to a minimum in $t_a$ for an intermediate $n$ value. Above argument can easily be generalized for step addition of stimulus as well.

The occurrence of $t_a$ minimum with $n$ can be explained even without using any specific expression for $z(t)$. The extremal time $t_a$ is the point from which activity variation reverses its trend. Because of $e^F$ ratio between the forward and reverse transition of activity switching, one transition is greatly favored over the other when $F$ undergoes sudden change at $t=0$. Starting from here, using the same line of reasoning as above, one can show that $t_a$ has a minimum with $n$. The specific relationship satisfied by activity and free energy at $t_a$ is not crucial to explain this minimum. Therefore, even if activity switching rates are chosen differently from what we have used in our present model, similar effect can be found.

\begin{figure}[h!]
\includegraphics[scale=0.7]{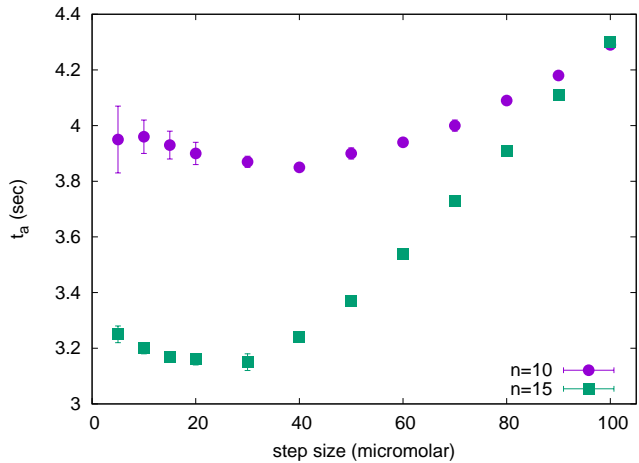}
\caption{$t_a$ shows a minimum with the size of the step stimulus for a fixed $n$. The circles present data for $n=10$ and the square symbols correspond to $n=15$. We have used a pre-stimulus attractant concentration of $[L]_0=200 \mu M$ here and varied $[L]_1$ within the range from $205 \mu M$ to $300 \mu M$. The optimum step size decreases with $n$. }
\label{fig:step}
\end{figure}

 Note that when a larger step size is used, then the change in $F$ is also larger. Then it follows from the above discussion that $t_a$ would then start increasing with $n$ even for relatively smaller $n$ values. This means the position of the minimum would shift leftward towards smaller $n$. This is consistent with our data in Fig. \ref{fig:tm} (left panel). One interesting outcome of this argument is that for a fixed $n$, if one just varies the step size, it should be possible to see a minimum in $t_a$. In other words, there exists an optimal step size for which extremal response is reached in the smallest possible time. Indeed our data in Fig. \ref{fig:step} verify this interesting prediction. This result can be directly tested in experiments. From the above argument we also expect that as $n$ increases, the optimum step size should be lower, and our data in Fig. \ref{fig:step} verify this trend.

\section{Extremal value of CW bias } \label{sec:cw}

\begin{figure}[h!]
\includegraphics[scale=0.7]{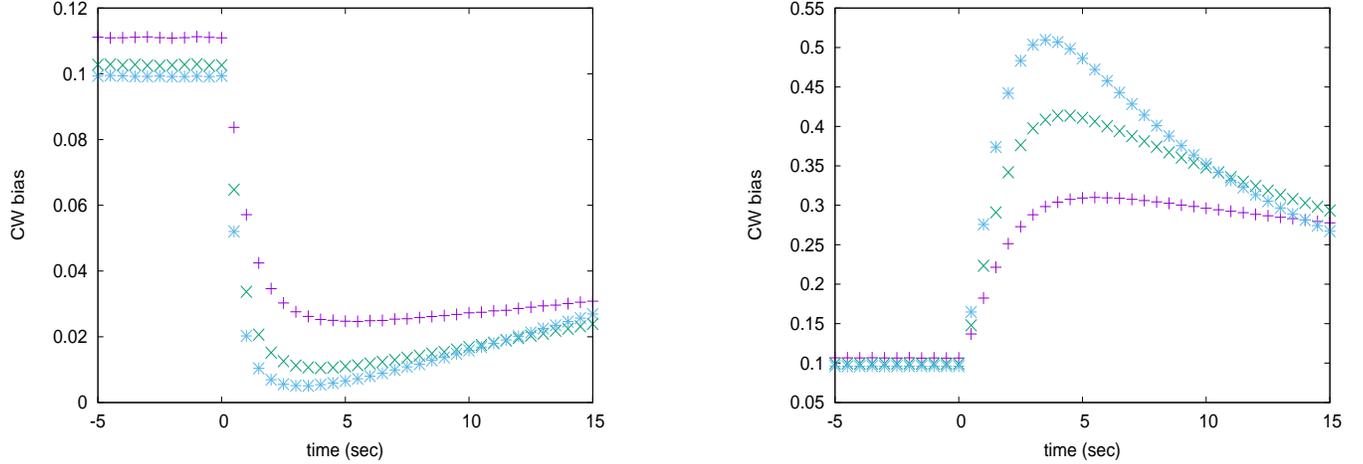}
\caption{Temporal variation of CW bias in response to addition (removal) of step stimulus. The left panel shows data for step addition where attractant level has been changes from $[L]_0=200 \mu M$ to $[L]_1=220 \mu M$ at time $t=0$. The right panel contains data $[L[_0=220 \mu M$ and $[L]_1=200 \mu M$. Here, we have shown data for $n=5,10,15$ with purple plus, green cross, blue star, respectively. }
\label{fig:cwb}
\end{figure}

\begin{figure}[h!]
\includegraphics[scale=0.7]{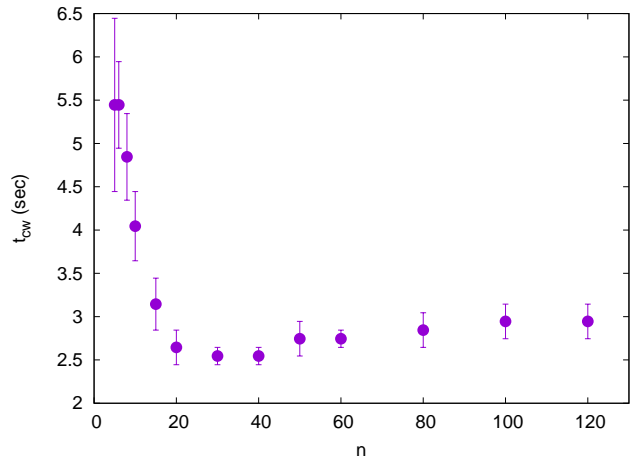}
\caption{$t_{cw}$ shows a minimum with $n$, for a fixed step size of the stimulus. Owing to large fluctuations in the CW bias data, the error-bars here are significantly larger. Here we have used $[L]_0=200 \mu M$ and $[L]_1=220 \mu M$.} \label{fig:cwm}
\end{figure}

Apart from activity, another experimentally relevant response function is the CW bias of the cell. Following application of a step stimulus, the variation of CW bias is recorded in experiments and the change in cell motility inferred from there \cite{min2012chemotactic}. In our simulations, we measure CW bias, defined as the probability to find the motors in the CW rotation mode. In Fig. \ref{fig:cwb} we show the temporal variation of CW bias for step addition and step removal of stimulus for different values of $n$. In the pre-stimulus state, CW bias remains constant with time. Immediately after the step stimulus is applied, CW bias starts changing rapidly with time, reaches an extremum value at time $t_{cw}$ beyond which its recovery starts. As found in the case of activity variation, the magnitude of the extremal response increases with $n$ for the same size of the step stimulus. From the left and right panels of Fig. \ref{fig:cwb} it can be seen that there is a significant asymmetry in the magnitude of extremal response for step addition and step removal. The extremal deviation from the pre-stimulus CW bias is much larger for the case of step removal than that of step addition. This is not surprising since the pre-stimulus value of CW bias is small and for step addition when CW bias decreases, its minimum value has to lie between the pre-stimulus value and zero. This restricts the magnitude of extremal response for step addition. However, for step removal, the CW bias can (in principle) increase all the way up to unity starting from a small pre-stimulus value. In other words, for an {\sl E. coli} cell in adapted state, the duration of CW rotation of the flagellar motors is much smaller ($\sim 0.2s$) compared to CCW rotational state duration ($\sim 1s$) which means the adapted value of CW bias is quite small. This makes the extremal response for step addition much smaller than step removal. In Fig. \ref{fig:cwasym} in the Appendix this effect is shown clearly. Our data in Fig. \ref{fig:cwm} also show that even for $t_{cw}$ there is an optimum $n$ for which $t_{cw}$ is minimum, although the error-bars are larger in this case and the minimum is comparatively shallower than what we had found for $t_a$ in Fig. \ref{fig:step}.  

\subsection{Exact calculation for extremal CW bias}
\begin{figure}
\includegraphics[scale=0.7]{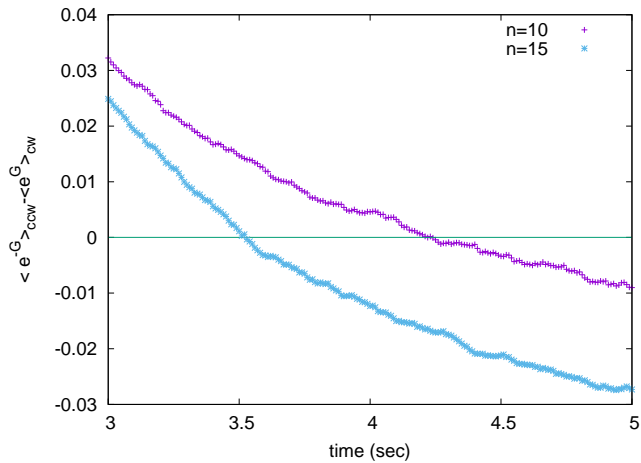}
\caption{Right hand side of Eq. \ref{eq:cwmin} vs time. We have presented data for $n=10,15$. These data are for $[L]_0 = 220 \mu M$ and $[L]_1 =200 \mu M$. The corresponding $t_{cw}$ values are $4.3 \pm 0.2$ sec for $n=10$ and $3.6 \pm 0.2 $ sec for $n=15$. The zero-crossing of the curves happens at times which match well with $t_{cw}$. }
\label{fig:cwdb}
\end{figure}

To analytically calculate the criterion for the extremal CW bias, we define $Q(cw,N_1,t)$ as the joint probability that the cell is found in CW-rotation mode with $N_1$ number of receptor clusters in the active state at time $t$. Similarly, $Q(ccw,N_1,t)$ can be defined. The master equation has the form
\bea
\nonumber
\partial_t Q(cw,N_1,t) &=& \omega e^{-G(N_1)} Q (ccw,N_1,t) - \omega e^{G(N_1)} Q(cw,N_1,t) \\ 
\nonumber
 &+& \Gamma_{1 \to 0}(t)\left [(N_1+1)Q(cw,N_1+1,t) -N_1 Q(cw,N_1,t)\right ] \\
 &+& \Gamma_{0 \to 1}(t)\left [ (N-N_1+1)Q(cw,N_1-1,t) -(N-N_1)Q(cw,N_1,t) \right ].
\label{eq:cwt}
\eea
Here, $\omega e^{\pm G}$ are the activity dependent switching rates between the CCW and CW modes described in Appendix \ref{app:model} and the rates $\Gamma$ have the same definition as in Eq. \ref{eq:pn1}. Summing both sides of Eq. \ref{eq:cwt} over $N_1$ causes cancellation of all terms containing $\Gamma$ and we are left with
\be
\partial_t {\cal Q}(cw,t) = \omega \left ( \langle e^{-G} \rangle_{ccw} - \langle e^G \rangle_{cw} \right)
\label{eq:cwmin}
\ee 
where ${\cal Q}(cw,t)$ is the CW bias (defined as the probability to find the cell in the CW mode at time $t$) and the right hand side contains averages performed over CCW mode and CW mode: $\langle e^G \rangle_{cw} = \sum_{N_1} e^{G(N_1)} Q(cw,N_1,t) $, etc. The left hand side of Eq. \ref{eq:cwmin} is zero at $t=t_{cw}$ when the CW bias reaches an extremum. In our simulations, we measure the averages $\langle e^G \rangle_{cw}$ and $\langle e^{-G} \rangle_{ccw}$ as a function of time. Physically, $\langle e^G \rangle_{cw}$ is the total transition flux from CW to CCW state at time $t$, and $\langle e^{-G} \rangle_{ccw}$ is the flux of reverse transition. The difference between these two fluxes gives rate of change of tumbling bias, as shown in Eq. \ref{eq:cwmin}. We plot this flux difference in  Fig. \ref{fig:cwdb} and our data clearly show that at $t=t_{cw}$ we indeed have 
\be 
\langle e^{-G} \rangle_{ccw} = \langle e^G \rangle_{cw}.
\label{eq:cwadapt}
\ee
This equality is expected to be satisfied when the system is in an adapted state. We have verified in our simulations (data not shown here) Eq. \ref{eq:cwadapt} holds in the pre-stimulus state, and long time after the stimulus is applied. Our data in Fig. \ref{fig:cwdb} shows at short time $t=t_{cw}$ also this equality is satisfied, as predicted by our simple calculation.

\section{Conclusions}

In this work we have studied short time response of an  {\sl E. coli} cell after application of a step stimulus. In contrast to many experimental and theoretical studies, where long time limit of the post-stimulus recovery is considered, our work highlights the importance of understanding the short time response of the cell. For example, many experiments measure the time-scale over which receptor activity or motor bias show half-way recovery  \cite{min2012chemotactic} \cite{keegstra2017phenotypic} \cite{frankel2014adaptability} but we show here it is important to take into account the variation of short time maximal response even before the recovery starts, and useful insights about the adaptation kinetics can be obtained from there. For example, the asymmetry in halfway recovery time-scale for step addition and step removal of stimuli has already been reported in literature \cite{min2012chemotactic}. We show here even the short time response of the cell shows this asymmetry. Our exact calculation for the extremal response at short times match well with our numerical simulations.

Many of our conclusions can be verified in experiments. The presence of an optimum step size of the stimulus which yields fastest extremal response is quite intriguing and can be directly tested using a tethered assay and FRET based setup \cite{colin2017multiple} \cite{keegstra2017phenotypic}. Experimentally it has also been possible to vary the strength of the cooperative interaction among the receptors which ultimately controls the size of the receptor clusters \cite{colin2017multiple}. Therefore it should be possible to check if there is an optimum interaction strength at which activity or CW bias hit their extremal values in the shortest possible time. Also, from the extremal activity value, one can indirectly estimate the methylation levels even when the system is far from adaptation. Note that it has not been possible till now to experimentally measure the receptor methylation levels in vivo and it can only be calculated indirectly from the knowledge of activity levels in the adapted state. Our results make it possible to get an idea of typical methylation levels of the receptors even when the post-stimulus activity is at its extremum.

How general are our results? Our main results remain valid irrespective of the kinetic details of the methylating or demethylating enzymes that regulate the adaptation module of the signaling network. Our conclusions are robust and should hold even for a model different from the one we have used here. We believe our research would open up new avenues of theoretical and experimental investigation of short time response of {\sl E. coli} chemotaxis.

\section{Acknowledgements} 
SC acknowledges financial support from the Science and Engineering Research Board, India (Grant No: MTR/2019/000946).

\appendix
\newpage
\section{Additional details of the model}
\label{app:model}

As mentioned in the main paper, the phosphorylation ability of the receptors defines its activity state. The receptors form a complex with the cytoplasmic protein CheA through the adaptor protein CheW. When attractant molecules bind to the receptors, the receptor goes to an inactive state and the autophosphorylation of CheA is suppressed. In absence of this binding, the receptor stays active and enhances autophosphorylation of CheA. In the phosphorylated state, CheA-P donates its phosphate group to demethylation enzyme CheB and response regulator CheY. In Fig. \ref{fig:model} we present a schematic diagram of the signaling network.

In our model, there are three major parts: (a) activity switching of the receptor clusters, (b) binding-unbinding dynamics of the enzyme molecules to the receptor dimers and (de)methylation of the receptor dimers by the bound enzymes, (c) switching of motor bias between CCW and CW rotation modes. A detailed description of each part is given below.

(a) We denote the activity state of the $i-$th receptor cluster containing $n$ trimers of dimers by the variable $a_i$, which can take two values. $a_i=1$ denotes an active state and $a_i=0$ denotes inactive state. The free energy difference $F$ between these two states is given by Eq. 2 of the main paper and the probability to find a receptor cluster in active state is $[1+\exp(F)]^{-1}$ when the system is fully adapted. From $a_i=0$ state the receptor cluster switches to $a_i=1$ state with the rate $\dfrac{w_a}{1+\exp(F)}$ and the reverse transition happens with a rate $\dfrac{w_a \exp(F)}{1+\exp(F)}$ \cite{colin2017multiple}. The choice of these rates is based on local detailed balance \cite{colin2017multiple}. The parameter $w_a$ is the characteristic time-scale of the transition. Here we have used $w_a=0.75 s^{-1}$ which is a little higher than the value $0.25 s^{-1}$ used in literature to explain the experimental data \cite{colin2017multiple, parkinson2015signaling} but we have verified (data not shown here) that this difference does not affect our conclusions \cite{shobhan} \cite{mandal2022effect}.

(b) Let $N_R$ and $N_B$ denote the total number of CheR and CheB molecules in the cell. In the unbound state these enzyme molecules reside in the cell cytoplasm and can bind to a receptor dimer if and only if no other enzyme molecules are bound to it. Each receptor dimer has a tether site and a modification site at which the binding can take place  \cite{wu1996receptor} \cite{feng1999enhanced} \cite{pontius2013adaptation}. Because of very few number of enzyme molecules compared to the large number of receptors in the cell, these binding events are often slow. We only consider tether binding in our model because it is relatively faster than the binding at modification sites \cite{schulmeister2008protein, pontius2013adaptation}. An unbound CheR molecule can bind to a dimer with a rate $w_r$ and once bound, it can raise the methylation level of that dimer with a rate $k_r$, provided the dimer has methylation level $< 8$ and it belongs to an inactive cluster of receptors. A bound CheR can unbind from the dimer with rate $w_u$ and then it can either reattach with another free dimer within the same cluster, or return to the cell cytoplasm. An unbound CheB molecule can undergo phosphorylation by an active receptor with a rate $w_p$ and its dephosphorylation happens with a rate $w_{dp}$. A CheB-P can bind, unbind, rebind in a similar way as described above for CheR. A CheB-P bound to an active dimer can demethylate it with rate $k_B$ provided its methylation level is $>0$. The possibility of rebinding of an enzyme within the same receptor cluster allows methylation modification of multiple neighboring dimers by the same enzyme molecule, which helps in maintaining robust adaptation \cite{endres2006precise, hansen2008chemotaxis, li2005adaptational, levin2002binding}.

(c) If $a$ denotes the fraction of active receptor clusters in the cell, then the phosphorylated fraction of CheY molecules, defined as $Y_P = \dfrac{[\text{CheY-P}]}{[\text{CheY}]}$ follows the rate equation \cite{flores2012signaling}
\be 
\frac{dY_P}{dt} = K_Y a (1-Y_P) - K_Z Y_P
\ee
where the parameters $K_Y$ and $K_Z$ are rates for phosphorylation and dephosphorylation, respectively. In our simulations we have directly used $Y_P = \dfrac{a}{a+K_Z/K_Y}$. The rotational bias of the flagellar motors are decided by $Y_P$. If a motor is in CCW state, it can switch to CW mode with a rate $\omega \exp (-G)$ and the reverse transition happens with the rate $\omega \exp(G)$. Here, $G = \Delta_1 - \dfrac{\Delta_2}{1+Y_0/Y_P}$.

\begin{table}[h!]
\caption{List of parameter values used in simulations}
\begin{tabular}{|l|l|l|l|}
\hline
\textbf{Symbol} & \hspace{30mm} \textbf{Description} & \textbf{Value} & \textbf{References} \\
\hline
$N_{dim}$ & Total number of receptor dimers & $7200$ & \cite{pontius2013adaptation,li2004cellular} \\
\hline
$N_R$  & Total number of CheR  protein molecules &  $140$ & \cite{pontius2013adaptation,li2004cellular}\\
\hline
$N_B$  & Total number of CheB  protein molecules &  $240$ & \cite{pontius2013adaptation,li2004cellular} \\
\hline
$K_{min}$ & Minimum concentration receptor can sense &  $18$ $\mu M$ & \cite{jiang2010quantitative}, \cite{flores2012signaling} \\
\hline
$K_{max}$ & Maximum concentration receptor can sense &  $3000$ $\mu M$ & \cite{flores2012signaling,jiang2010quantitative} \\
\hline
$w_{a}$  & Flipping rate of activity &  $0.75$ $s^{-1}$ & Present study\\
\hline
$\omega$  & Switching frequency of motor &  $1.3$ $s^{-1}$ & \cite{sneddon2012stochastic} \\
\hline
 $\Delta_{1}$  & Nondimensional constant regulating motor switching  &  $10$ & \cite{sneddon2012stochastic} \\
\hline
$\Delta_{2}$  & Nondimensional constant regulating motor switching &  $20$ & \cite{sneddon2012stochastic}  \\
\hline
$Y_{0}$ & Adopted value of the fraction of CheY-P protein &  $0.34$ & \cite{sneddon2012stochastic} \\
\hline 
$K_{Y}$ & Phosphorylation rate of CheY molecule &  $1.7$ $s^{-1}$ & \cite{flores2012signaling,tu2008modeling} \\
\hline
$K_{Z}$ & Dephosphorylation rate of CheY molecule &  $2$ $s^{-1}$ & \cite{flores2012signaling,tu2008modeling}\\
\hline
$w_{r}$ & Binding rate of bulk CheR to tether site of an unoccupied dimer &  $0.068$ $s^{-1}$ & \cite{pontius2013adaptation,schulmeister2008protein} \\
\hline
$w_{b}$ & Binding rate of bulk CheB-P to tether site of an unoccupied dimer &  $0.061$ $s^{-1}$ & \cite{pontius2013adaptation,schulmeister2008protein} \\
\hline
$w_{u}$ & Unbinding rate of bound CheR and CheB-P &  $5$ $s^{-1}$ & \cite{pontius2013adaptation,schulmeister2008protein} \\	
\hline	
$k_{r}$ & Methylation rate of bound CheR & $2.7$ $s^{-1}$ & \cite{pontius2013adaptation,schulmeister2008protein}\\ 	
\hline
$k_{b}$ & Demethylation rate of bound CheB-P &  $3$ $s^{-1}$ & \cite{pontius2013adaptation,schulmeister2008protein}\\
\hline	
$w_{p}$ & CheB phosphorylation rate &  $3$ $s^{-1}$ & \cite{pontius2013adaptation,stewart2000rapid} \\
\hline
$w_{dp}$ & CheB-P dephosphorylation rate &  $0.37$ $s^{-1}$ & \cite{pontius2013adaptation}\\	
\hline	
$dt$ & Time step &  $0.01$ $s$ & Present study\\
\hline
\end{tabular}
\label{table}
\end{table}

\newpage
\section{Adapted activity decreases with $n$}
\begin{figure}[h]
\includegraphics[scale=0.7]{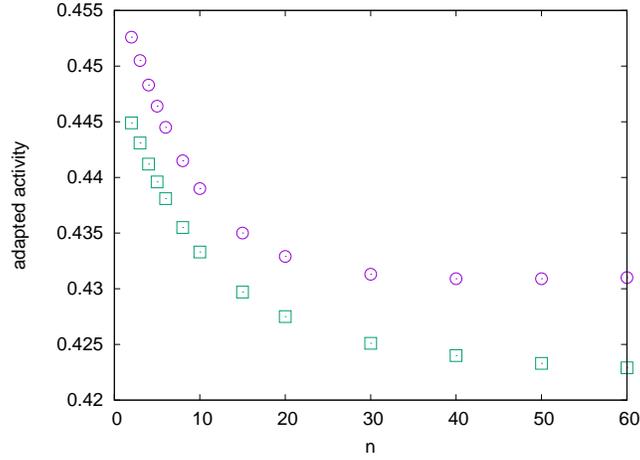}
\caption{Variation of pre-stimulus adapted activity with cluster size $n$. The upper curve corresponds to $[L]_0 = 200 \mu M$ and the lower curve is for $[L]_0=250 \mu M$. In both cases the adapted activity decreases weakly with $n$.}
\label{fig:a0}
\end{figure}

In Fig. \ref{fig:a0} we plot the variation of pre-stimulus adapted activity with $n$ for different $[L]_0$ values. Our data show that activity decreases with $n$ and $[L]_0$. For a fixed $[L]_0$ the drop is relatively sharper for small $n$ and tends to flatten out for large $n$. 

\newpage
\section{Asymmetric response for step addition and step removal}
\begin{figure}[h!]
\includegraphics[scale=0.7]{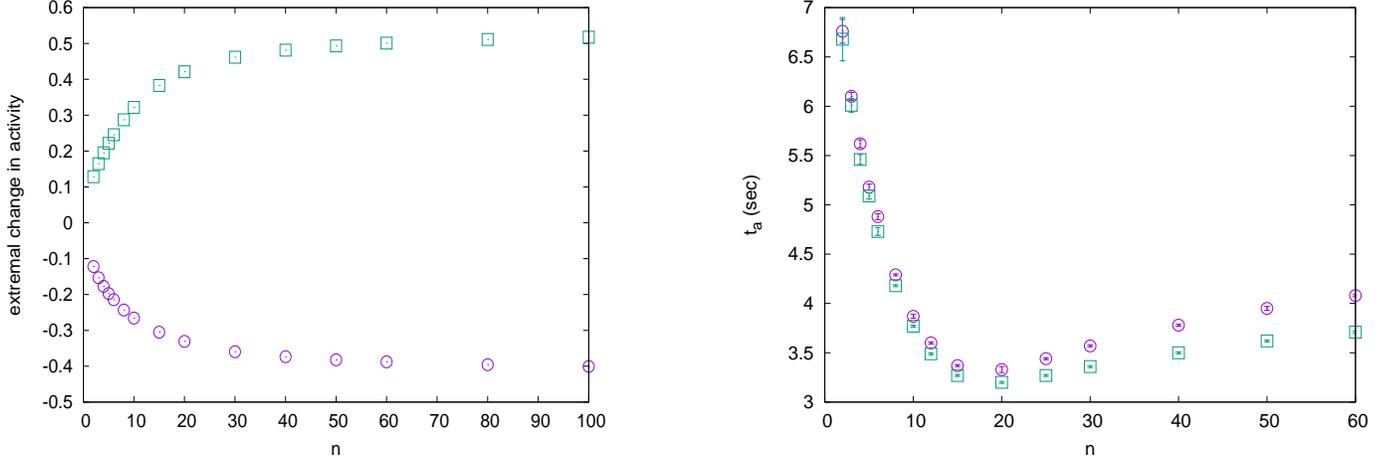}
\caption{Short time response shows asymmetry between step addition (purple circles) and step removal (green squares). Left panel: The difference between extremal activity level and pre-stimulus level shows opposite sign and different magnitudes when step up and step down stimuli are applied. The magnitude is higher for step removal. Right panel: $t_a$ values are higher for step addition case. For step up stimulus the attractant level is changed from $200 \mu M$ to $250 \mu M$ and for step down case, the reverse change is considered. The error-bars are shown which are smaller than the symbol size. }
\label{fig:asym}
\end{figure}

To demonstrate the asymmetric extremal response between step addition and step removal of stimulus, we measure the difference between extremal activity value $\langle a(t_a) \rangle$ and the pre-stimulus activity level $\langle a_0 \rangle$. In Fig. \ref{fig:asym} left panel we plot this difference as a function of $n$ for step addition (when the attractant level is changed from $200 \mu M$ to $250 \mu M$) and step removal (from $250 \mu M$ to $200 \mu M$). We find that the extremal response does not only change its sign for these two protocols, but the magnitude is also distinctly higher for the case of step removal. Right panel of Fig. \ref{fig:asym} show the corresponding $t_a$ values. We find $t_a$ is larger for the case of step addition.

\begin{figure}[h!]
\includegraphics[scale=0.7]{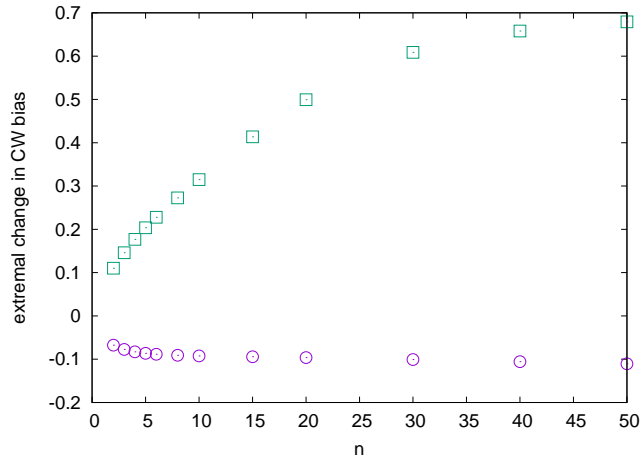}
\caption{The extremal deviation in CW bias from the pre-stimulus level plotted as a function of receptor cluster size. The square symbols represent the data for step addition when the attractant level is changed from $200 \mu M$ to $220 \mu M$, while the circles show the data for step removal when the attractant level is changed from $220 \mu M$ to $200 \mu M$. The extremal deviation is much larger for step removal. }
\label{fig:cwasym}
\end{figure}

For CW bias the asymmetry is even more pronounced. In Fig. \ref{fig:cwasym} we plot the extremal deviation of CW bias from its pre-stimulus value and these data show that the response is much larger for step-removal of attractant. As explained in Sec. \ref{sec:cw} of the main paper this asymmetry is due to small value of pre-stimulus CW bias which severely restricts the extremal deviation for step addition.

\newpage
\section{Extremal change in activity vs step size}
\begin{figure}[h!]
\includegraphics[scale=0.7]{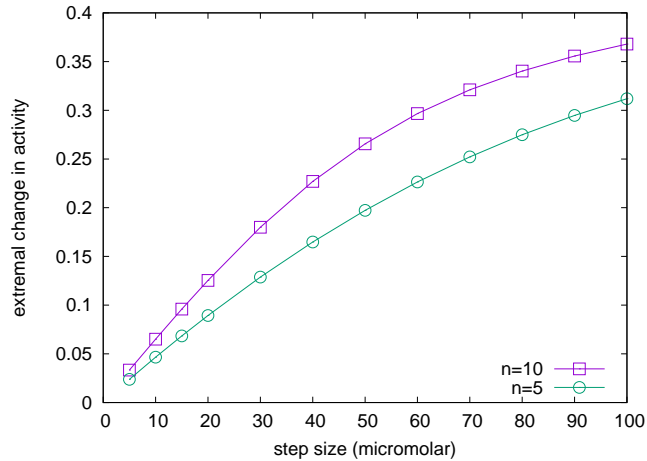}
\caption{ Extremal change in activity increases linearly with step size of the stimulus for small steps. For large steps the variation is slower than linear. The upper curve corresponds to $n=10$ and the lower curve is for $n=5$. As expected, deviation from linearity starts sooner for larger $n$. }
\label{fig:delact}
\end{figure}

We have already mentioned in the main text that we consider only a limited range of stimulus step size $[L]_1-[L]_0$ and receptor cluster size $n$ such that the extremal value of activity stays strictly above $0$ and strictly below $1$. This makes sure that $t_a$ is well-defined. Within this range, we find the difference between extremal activity and pre-stimulus activity increases linearly with step size when the steps are small. This is expected from linear response theory. As the step size increases, the linear growth slows down and shows the effect of saturation. We present our data in Fig. \ref{fig:delact}. 


\end{document}